# Development of a Three-Dimensional Multiscale Agent-Based Tumor Model:

## Simulating Gene-Protein Interaction Profiles, Cell Phenotypes & Multicellular Patterns in Brain Cancer

**Le Zhang [1], Chaitanya A. Athale [2] and Thomas S. Deisboeck [1*]**

[1] Complex Biosystems Modeling Laboratory, Harvard-MIT (HST) Athinoula A. Martinos Center for Biomedical Imaging, Massachusetts General Hospital, Charlestown, MA 02129, USA; [2] Cell Biology and Biophysics Department, European Molecular Biology Laboratory (EMBL), Heidelberg, D-69117, Germany.

**Running Title:**    Development of a 3D Multiscale Agent-Based Tumor Model

**\*Corresponding Author:**

Thomas S. Deisboeck, M.D.
Complex Biosystems Modeling Laboratory
Harvard-MIT (HST) Athinoula A. Martinos Center for Biomedical Imaging
Massachusetts General Hospital-East, 2301
Bldg. 149, 13th Street
Charlestown, MA 02129
Tel: 617-724-1845
Fax: 617-726-7422
Email: deisboec@helix.mgh.harvard.edu

## ABSTRACT

Experimental evidence suggests that epidermal growth factor receptor (EGFR)-mediated activation of the signaling protein phospholipase C $\gamma$ plays a critical role in a cancer cell's phenotypic decision to either proliferate or to migrate at a given point in time. Here, we present a novel three-dimensional multiscale agent-based model to simulate this cellular decision process in the context of a virtual brain tumor. Each tumor cell is equipped with an EGFR gene-protein interaction network module that also connects to a simplified cell cycle description. The simulation results show that over time proliferative and migratory cell populations not only oscillate but also directly impact the spatio-temporal expansion patterns of the entire cancer system. The percentage change in the concentration of the sub-cellular interaction network's molecular components fluctuates, and, for the 'proliferation-to-migration' switch we find that the phenotype triggering molecular profile to some degree varies as the tumor system grows and the microenvironment changes. We discuss potential implications of these findings for experimental and clinical cancer research.

# 1. INTRODUCTION

Malignant brain tumors such as glioblastoma exhibit complex growth patterns. Interestingly, experimental observations of such high-grade gliomas suggest that at the same point in time, migrating tumor cells do not proliferate and conversely proliferating ones do not migrate. While this led Giese et al. (1996) to propose the intriguing concept of "dichotomy" in gliomas, the exact molecular mechanism governing this reversible switch has not yet been clearly established and, moreover, the impact any such molecular event potentially has beyond the scale of a single cancer cell remains to be properly evaluated. In this situation, *in silico* modeling can help by integrating data and yielding experimentally testable hypotheses. In this context, it is noteworthy that the epidermal growth factor (EGF) receptor (EGFR) pathway has been shown to be involved at various steps in tumorigenesis, also in gliomas (Chicoine and Silbergeld, 1997), and its role for the phenotypic switch has already been suggested in the case of breast cancer (Dittmar et al., 2002). As a starting point, in an effort to simulate this phenotypic 'switch' behavior, we have therefore integrated a cell cycle module taken from the literature (Alacon et al., 2004; Tyson and Novak, 2001) into our previously developed EGFR gene-protein interaction network model (Athale et al., 2005). In our new model now, each cell utilizes the value state of its molecular network to 'decide' its microscopic phenotype, i.e., migration, proliferation, quiescence, or apoptosis - at every point in time. On the micro-macroscopic level, a fixed three-dimensional lattice is employed to represent a virtual block of brain tissue, while in the molecular environment, the phenotypic behavior of a cell is determined by the dynamical changes in the concentrations of the interacting molecular species both inside and around the tumor cell. The result is a closer step towards a comprehensive multiscale model of a malignant brain tumor that not only can forecast the overall tumor growth dynamics but also monitor the dynamical changes within each cell's molecular network and the profiles, respectively, that trigger the phenotypic switch. In the following section we briefly review relevant works.

## 2. PREVIOUS WORKS

The EGF/EGFR cell signaling system has been studied extensively, both in experimental and theoretical works. For instance, Starbuck and Lauffenburger (1992) suggested a mathematical model for receptor-mediated cell uptake and processing of EGF. This model simulates the mitogenic signal generated by EGF/EGFR binding to the cell surface via stimulation of receptor tyrosine kinase activity. In addition, Chen et al. (1996) revealed a possible role for a phospholipase C dependent feedback mechanism that attenuates EGF-induced mitogenesis. Further, the model of Schoeberl et al. (2002) offered an integrated quantitative dynamic and topological representation of intracellular signal networks, based on known components of EGF receptor signaling pathways. Lastly, employing a hybrid modeling approach, Sander and Deisboeck (2002) argued that both strong heterotype chemotaxis and strong homotype chemoattraction, such as through the EGF analogue transforming growth factor alpha (TGF-α), are required for branch formation within the invasive zone of microscopic brain tumors. And indeed, Wang et al. (2005) demonstrated that chemotactic cell migration in response to EGF are correlated with invasion, intravasation and metastasis in animal models of breast cancer.

More recently, so-called multi-scale modeling platforms that span several biological levels of interest drew attention, because of their potential to integrate molecular and multicellular experimental data. For instance, in previous works from our laboratory (Mansury et al., 2002; Mansury and Deisboeck, 2003; Mansury and Deisboeck, 2004a, b), we concentrated on bridging the macroscopic, microscopic and molecular tumor scales. Specifically, Mansury and Deisboeck (2003) proposed a two-dimensional agent-based model in which the spatio-temporal expansion of malignant brain tumor cells is guided by environment heterogeneities in mechanical confinement, toxic metabolites and nutrient sources to gain more insight into the systemic effect of such cellular chemotactic search precision modulations. With this model they continued to investigate the relationship between rapid growth and extensive tissue infiltration (Mansury and Deisboeck, 2004a). Moreover, by calibrating the expression of Tenascin C and PCNA using experimental brain tumor data for the migratory phenotype while generating the gene expression for proliferating cells as the output, numerical result from this model (Mansury and Deisboeck, 2004b) confirmed that among the migratory phenotype the expression of Tenascin C is indeed consistently higher, while they reveal the reverse for the proliferating tumor cells, which exhibit consistently higher expression of the proliferating cell nuclear antigen (PCNA) gene. Athale et al. (2005) extended this agent-based, primarily 'micro-macro' framework down to an even more enriched sub-cellular level, thus developing a multiscale cancer model that allows monitoring the percolation of a molecular perturbation throughout the emergent multi-cellular system. Particularly, this model introduced a simplified EGFR pathway as a signal processing module that encodes the switch between the cell's microscopic phenotypes of proliferation and migration. The results showed further *in silico* evidence that behavioral decisions on the single cell level impact the spatial dynamics of the entire cancerous system. Furthermore, the simulation results yielded intriguing experimentally testable hypotheses such as spatial cytosolic polarization of $PLC_\gamma$ towards an extrinsic chemotactic gradient (Devreotes and Janetopoulos, 2003). While this work already implicitly acknowledged the existence of a cell cycle, it however lacked a detailed representation of the cell cycle. Based on the works by Tyson & Nowak (2001) who represented eukaryotic molecular mechanisms as sets of nonlinear ordinary differential equations and used standard analytical and numerical methods to study their solutions, Alarcon et al. (2004) applied a revised version of their model to the case of cancer cells under hypoxic conditions. While Alarcon et al (2004) left hypoxia tension constant, in our study here, hypoxia tension is considered a dynamic external condition. Therefore, we first modified their cell cycle module to be able to correlate it with the location of the cell and then integrated it into our previously developed *multiscale agent-based model* (Athale et al., 2005). The next section details the setup of the model.

## 3. MATHEMATICAL MODEL

Our multi-scale model incorporates both macro-microscopic and molecular environments. In the following sections, we will illustrate the characteristics of these environments, proceeding in a top-down manner.

### 3.1. Macro-microscopic environment

We first create a three dimensional rectangular lattice that consists of a grid with $100 \times 100 \times 100$ points in size representing a block of virtual brain tissue. Each lattice site will be assigned a value of $TGF_\alpha$ ($X_1$), of glucose ($X_{14}$), and oxygen tension ($k_{44}$) representing these external chemical cues by normal distribution. To display the effect of chemotaxis, the three dimensional lattice is divided into four cubes, depicted in **Figure 1**.

**Figure 1.**

The levels of these distributions are weighted by the distance, $d_{ijk}$, of a given cell from the center of cube 4, computed by the previously reported L-infinity metric of measurement (Mansury et al., 2002). The center of cube 4 is assigned the highest glucose and $TGF_\alpha$ concentrations as well as oxygen tension value, hence rendering it the most "attractive" for the chemotactically acting tumor cells. Note that each grid point can be occupied by only one cell at each time step. The chemotaxis distributions of $TGF_\alpha$, glucose and oxygen tension are described by these underlying equations:

$$X_1^{ijk} = T_m \exp(-2d_{ijk}^2 / \sigma_t^2) \tag{1a}$$

$$X_{14}^{ijk} = G_a + (G_m - G_a)\exp(-2d_{ijk}^2 / \sigma_g^2) \tag{1b}$$

$$k_{44}^{ijk} = k_a + (k_m - k_a)\exp(-2d_{ijk}^2 / \sigma_o^2) \tag{1c}$$

In **Eq. 1a**, $T_m$ stands for the maximum $TGF_\alpha$ concentration in the tumor (Moskal et al., 1995), $\sigma_t$ is the parameter that controls the dispersion of the $TGF_\alpha$ level. In **Eq. 1b**, $G_a$ is the minimum blood glucose level while $G_m$ stands for the maximum concentration of glucose in blood (Freyer and Sutherland, 1986), with $\sigma_g$ being the parameter controlling the dispersion of glucose. Likewise, in **Eq. 1c**, $k_a$ is the minimum oxygen tension and $k_m$ represents the maximum oxygen tension (Alarcon et al., 2004), $\sigma_o$ is the parameter controlling the dispersion of the oxygen tension level. Note that high oxygen tension equals a low level of hypoxia and vice-a-visa.

To begin with, five hundred cancer cells are initialized at the center of the lattice [50,50,50] and a replenished nutrient point source, representing a blood vessel, is set in the center of cube 4. When the first cancer cell reaches this location, the simulation is terminated. Glucose and external $TGF_\alpha$ ($TGF_\alpha\_ex$) concentrations are the two major chemoattractive cues in this macro-microscopic environment. Glucose ($X_{14}$) continues to diffuse throughout the three dimensional lattice with a fixed rate and only the location that harbors the peak concentration is replenished at each time step. Furthermore, as a nutrient it is continuously taken up by cells to maintain their metabolism. In **Eq. 2a**, $t$ represents the time step while $ijk$ stands for the position on the three dimensional lattice and $r_n$ is the cell's glucose uptake coefficient (Mansury et al. 2002). $D_1$ is the diffusion coefficient of the glucose (Sander and Deisboeck, 2002).

$$X_{14}^t = X_{14}^{t-1} - r_n \tag{2a}$$

$$\frac{\partial X_{14}^{ijk}}{\partial t} = D_1 \nabla^2 X_{14}^{ijk}, t = 1,2,3... \tag{2b}$$

$TGF_\alpha$, is an autocrine produced hormone which can, in addition, act in a paracrine fashion as well as juxtacrine manner (Shvartsman et al., 2001) through triggering, as EGF analogue, the EGF-receptor pathway. Thus, cells not only can take up their own $TGF_\alpha$ but also that secreted by bystander cells. Here, $TGF_\alpha$ ( $X_1$ ) degrades and diffuses to its neighborhood at each time step. Furthermore, (aside from the cells' autocrine secretion) its 'external' replenishment is again restricted only to the site of the virtual blood vessel in cube 4 (compare with **Figure 1.**). It follows that

$$X_1^t = X_1^{t-1} + S_T \tag{2c}$$

$$\frac{\partial X_1^{ijk}}{\partial t} = D_2 \nabla^2 X_1^{ijk}, t = 1,2,3... \tag{2d}$$

where $D_2$ is the $TGF_\alpha$ diffusion coefficient (Thorne et al., 2004) and $S_T$ is the $TGF_\alpha$ secretion rate (Forsten and Lauffenburger, 1992).

Based on the experimentally determined diffusion distance of oxygen in tissue, we have limited it to a (rescaled) $100 \mu m$ distance from the location of the blood vessel (Carmeliet and Jain, 2000). Thus the oxygen diffusion process can be described with the following equation:

$$\frac{\partial k_{44}^{ijk}}{\partial t} = D_o \nabla^2 k_{44}^{ijk}, t = 1,2,3...$$ (2e)

where $D_o$ stands for the diffusion coefficient of oxygen (Carmeliet and Jain, 2000). The values of the aforementioned coefficients are listed in **Tables 1-4**.

**Table 1.-4.**

## 3.2. Molecular environment

Turning now to the molecular environment which is comprised of both, an EGFR gene-protein interaction network and a cell cycle subsystem. The EGFR gene-protein interaction network is designed to simulate how the cell processes its phenotype decision with regards to proliferation and migration whereas the cell cycle is added to complement the proliferation process explicitly.

### 3.2.1. EGFR gene-protein interaction network

As discussed in detail in Athale et al. (2005), induced by the state of its regulatory EGFR gene-protein interaction network (and its microenvironmental cues) a given cell will, at any point in time, choose its phenotypic trait such as migration, proliferation or quiescence or turn apoptotic. All molecular species and the coefficients of this network are listed in **Tables 1** and **2**, respectively.

**Figure 2.**

In brief, as displayed in **Figure 2** each agent or virtual tumor cell has four layers, i.e., the external space, the cell membrane, the cytoplasm and the nucleus. In the external space and membrane layers, there are glucose, $TGF_\alpha\_ex$, $EGFR\_s$, $TGF\alpha - EGFR\_s$ and $ppTGF_\alpha - EGFR\_s$ variables. $TGF_\alpha\_ex$ ($X_1$) binds to the receptor $EGFR\_s$ ($X_2$) and rapidly dimerizes to $2TGF_\alpha - EGFR\_s$ ($X_3$) (Starbuck and Lauffenburger, 1992) which in turn is then autophosphorylated to $2ppTGF_\alpha - EGFR\_s$ ($X_4$). The following equations represent these processes:

$$\frac{dX_1}{dt} = k_{-1} \cdot X_3 - k_1 \cdot X_1 \cdot X_2 + k_9 \cdot X_7 - k_{11} \cdot X_1 \tag{3}$$

$$\frac{dX_2}{dt} = k_{-1} \cdot X_3 - k_1 \cdot X_1 \cdot X_2 + k_8 \cdot X_6 - k_{-8} \cdot X_2 \tag{4}$$

$$\frac{dX_3}{dt} = 2 \cdot k_1 \cdot X_1 \cdot X_2 - 2 \cdot k_{-1} \cdot X_3 - k_2 \cdot X_3[1 + w_g X_{13}]$$
$$- k_3 \cdot X_3 + k_{-2} X_4 + \frac{V_{M2} . X_{11}}{K_{M2} + X_{11}} X_4 \tag{5}$$

$$\frac{dX_4}{dt} = k_2 \cdot X_3[1 + w_g X_{13}] - k_{-2} \cdot X_4 - k_4 X_4 - \frac{V_{M2} . X_{11}}{K_{M2} + X_{11}} X_4 \tag{6}$$

Once internalized, the cytoplasmatic $TGF_\alpha - EGFR$ complex ($X_5$) dissociates reversibly to cytoplasmic $TGF_\alpha$ ($X_6$) and EGFR ($X_7$), denoted by

$$\frac{dX_5}{dt} = k_3 \cdot X_3 + k_4 \cdot X_4 + 2 \cdot k_{-5} \cdot X_6 \cdot X_7 - 2 \cdot k_5 \cdot X_5 \tag{7}$$

$$\frac{dX_6}{dt} = k_5 \cdot X_5 - k_{-5} \cdot X_6 \cdot X_7 - k_8 \cdot X_6 + k_{-8} \cdot X_2$$
$$+ k_{12} \cdot X_8 - k_6 \cdot X_6 \tag{8}$$

$$\frac{dX_7}{dt} = k_5 \cdot X_5 - k_{-5} \cdot X_6 \cdot X_7 - k_9 \cdot X_7 + k_{15} \cdot X_9 - k_7 \cdot X_7 \tag{9}$$

There is evidence that increased internalization of $X_3$ and $X_4$ leads to down-regulation of EGFR RNA ($X_9$) expression and thus diminished protein content (Hamburger et al., 1991), whereas *EGFR* activation by ligand binding increases $TGF_\alpha$ RNA ($X_8$) synthesis. Both RNA species are being transcribed and translated at a constitutive rate (Maruno et al., 1991; Van der Valk et al., 1997) and both, protein and RNA are constantly degraded (Mader, 1988),

$$\frac{dX_8}{dt} = k_{13} \cdot X_{12} - k_{14} \cdot X_8 \tag{10}$$

$$\frac{dX_9}{dt} = k_{17} \cdot X_{12} - k_{16} \cdot X_9 + k_{18} \cdot X_4 \tag{11}$$

with $X_{12}$ being the pool of nucleotides. The increased phosphorylated $TGF_\alpha - EGFR$ complex accelerates the rate of transition from inactive $PLC_\gamma$ ($X_{10}$) to active $PLC_\gamma$ ($X_{11}$). This active $PLC_\gamma$ exhibits negative feedback inhibition of $X_4$ (Chen et al., 1994, 1996; Wells, 1999) and is represented by

$$\frac{dX_{10}}{dt} = k_{21} \cdot X_{11} - k_{20} \cdot (k_{29} - X_{11}) \cdot X_4 \tag{12}$$

$$\frac{dX_{11}}{dt} = k_{20} \cdot (k_{29} - X_{11}) \cdot X_4 - k_{21} \cdot X_{11} \tag{13}$$

$$\frac{dX_{12}}{dt} = k_{16} \cdot X_9 + k_{14} \cdot X_8 - k_{13} \cdot X_{12} - k_{17} \cdot X_{12} \tag{14}$$

The intracellular glucose concentration ($X_{13}$) increases through uptake (Noll et al., 2000) from the extracellular glucose pool ($X_{14}$) yet is depleted by both, $TGF_\alpha - EGFR$ phosphorylation (**Eq. 4**) (Hertel et al., 1986; Steinbach et al., 2004) and cell metabolism (**Eq. 1**), and is described as

$$\frac{dX_{13}}{dt} = k_{23} \cdot X_{14} - k_2 \cdot X_3 \cdot X_{13} - k_{28} \cdot X_{13} \tag{15}$$

### 3.2.2. Cell cycle

The central element of our simplified module here is the biological 'on-off' cell cycle switch put forward by Tyson and Novak (2001). Our network also accounts for the effect of hypoxia and protein p27 on cell division by incorporating the module previously developed by Alarcon et al. (2004). In their cell cycle module, the switching behavior arises from the antagonism between cdh1-APC complexes ($X_{15}$) and cyclin-CDK ($X_{16}$), with the mass of cell ($X_{17}$) triggering the aforementioned switch. Cell division occurs when $X_{15} < k_{30}$ and $X_{16} > k_{31}$, where $k_{30}$ and $k_{31}$ denote thresholds of cdh1-APC complexes and cyclin-CDK, respectively. From **Figure 2** we can deduce that protein p27 ($X_{18}$) is up-regulated under hypoxic conditions. **Eq. 19** is employed to show this relationship, i.e., if hypoxia is low, the cell will have less protein p27, implying a shorter cell cycle. Inversely, if hypoxia is high, more protein p27 is generated, resulting in an inhibition of the cell cycle. This inhibitory effect of protein p27 ($X_{18}$) on the cyclin-CDK ($X_{16}$) is incorporated in **Eq. 17** through an additional decay term that is proportional to the concentration of p27. Moreover, **Eq. 18** describes the value changes of the mass of the cell ($X_{17}$) during cell cycle which impacts p27 ($X_{18}$) via **Eq. 19**. Another feature of Alarcon et al.'s (2004) module is the effect of phosphorylated retinoblastoma protein, RB. That is, non-phosphorylated RB (RBNP ($X_{19}$)) is known to inhibit CDK ($X_{16}$) activity (Knudsen et al., 1999) while phosphorylated RB has no direct effect (Gardner et al., 2001). To depict this process, RBNP is incorporated into **Eq. 16** as a generic activator for cdh1-APC complexes ($X_{15}$), which is implicit to inhibit cyclin-CDK ($X_{16}$) by **Eq. 17**. Variables and coefficients of this cell cycle module are listed in **Table 3** and **Table 4**.

$$\frac{dX_{15}}{dt} = \frac{(1+k_{32}X_{19})(1-X_{15})}{k_{34}+1-X_{15}} - \frac{k_{33}X_{17}X_{15}X_{16}}{k_{35}+X_{15}} \tag{16}$$

$$\frac{dX_{16}}{dt} = k_{39} - (k_{36}+k_{37}X_{15}+k_{38}X_{18})X_{16} \tag{17}$$

$$\frac{dX_{17}}{dt} = k_{40}X_{17}(1-\frac{X_{17}}{k_{41}}) \tag{18}$$

$$\frac{dX_{18}}{dt} = k_{42}(1-\frac{X_{17}}{k_{41}}) - k_{43}\frac{k_{44}}{k_{44}+k_{45}}X_{18} \tag{19}$$

$$\frac{dX_{19}}{dt} = k_{46} - (k_{47}+k_{46}X_{16})X_{19} \tag{20}$$

We note that throughout the simulation the cell cycle time ranges between $25 \pm 6.2$ hrs and is thus in good agreement with data reported in the literature such as in Hegedus et al. (2000).

### 3.3. Cell phenotypes

### 3.3.1. Proliferation and migration

The signaling protein phospholipase $C_\gamma$, $PLC_\gamma$, is known to be involved in directional cell movement in response to EGF (Mouneimne et al., 2004) and prognostic relevance of $PLC_\gamma$ expression in patients with glioblastoma has already been reported by Mawrin et al. (2003). For our purposes here particularly noteworthy, Dittmar et al. (2002) demonstrated that $PLC_\gamma$ is activated transiently in cancer cells, that is to a greater extent during migration and more gradually in the proliferating mode. Implementing this concept, we adopt the threshold, $\sigma_{PLC}$, to decide if the cell should undergo migration or not. Each cell is therefore evaluated for its migratory potential ($MP$),

$$MP[X_{11}] = [\frac{dX_{11}}{dt}] \tag{21}$$

where $\frac{dX_{11}}{dt}$ is the change in concentration of $PLC_\gamma$ over time. If $MP$ is greater than the $\sigma_{PLC}$, the cell chooses to migrate, otherwise it proliferates or remains quiescent. If the cell decides to migrate, it searches for the best location in its vicinity to move to. The candidate 'best' locations are comprised of all the 'Von Neumann' neighborhood sites (Athale et al., 2005) of this cell. **Eq. 22** (Mansury and Deisboeck, 2003) describes how the cell chooses this most attractive location according to:

$$T_j = \psi \cdot L_j + (1-\psi)\varepsilon_j \tag{22}$$

where $T_j$ stands for the perceived attractiveness of location $j$, $L_j$ represents the correct, non-erroneous evaluation of location $j$ that will be defined further below, where $\varepsilon_j \sim N(\mu, \sigma^2)$ is an error term that is normally distributed with mean $\mu$ and variance $\sigma^2$. The parameter $\psi$ is positive between zero and one, $0 \leq \psi \leq 1$, and represents the extent of the *search precision*. For example, $\psi = 1$ represents a chemotactic search process operating with a 100 percent precision, i.e. tumor cells always evaluate the permissibility of a location without any processing error in the receptor network. By contrast, when $\psi = 0$, tumor cells perform a random-walk motion. However, if MP is less than $\sigma_{PLC}$ and $ppTGF\alpha - EGFR\_s$ is greater than $\sigma_{EGFR}$ the new cell will occupy one of its Von Neumann neighborhood sites with highest $T_j$ value (glucose concentration), otherwise it will become quiescent. Based on the results presented in Mansury and Deisboeck (2003) we chose here a search precision of 0.7 as this had been shown to ensure maximum spatio-temporal expansion of the virtual tumor system.

### 3.3.2. Quiescence and apoptosis

There are three possibilities that the cell enters the reversible quiescent state: (1) the cell is unable to find an unoccupied location to migrate or proliferate into; or (2) the migration potential ($MP$) is less than $\sigma_{PLC}$ and $ppTGF\alpha - EGFR\_s$ is less than $\sigma_{EGFR}$; and finally, (3) the glucose concentration around the cell ranges in between 16 mmol/L and 8 mmol/L. Note that if the on site glucose concentration diminishes even further, i.e. below 8 mmol/L, the cell turns apoptotic (Freyer and Sutherland, 1986).

## 4. RESULTS

Our code here is implemented in Java (Sun Microsystems, Inc., USA) and employs an agent-based modeling toolkit (http://cs.gmu.edu/~eclab/projects/mason/) that is combined with in-house developed classes for representing molecules, reactions and multi-receptors as a set of hierarchical objects. Running the simulation 10 times with different random normal distribution, $\sigma_g$, of glucose (**Eq. 1b**), the algorithm requires a total of 25 h 46 min of CPU time on a computer with an IBM Bladecenter machine (dual-processor 32-bit Xeons ranging from 2.8-3.2GHz w/2.5GB RAM) and Gigabit Ethernet. Each node runs Linux with a 2.6 kernel and Sun's J2EE version 1.5.

**Volumetric growth dynamics:** We measure the tumor system's [total] volume by counting the number of the lattice sites occupied by a tumor cell regardless of its phenotype, hence lumping together both proliferative and migratory driven expansion. **Figure 3** shows the increase in tumor system volume over time for the 10 simulation runs. The volume increase is not smooth, rather shows "jumps" or steps at distinct time points with a marked *acceleration* of the growth rate at later stages.

**Figure 3.**

**Macroscopic behavior:** Based on several of these stepping points reported in **Figure 3**, we display in **Figures 4.(a)-(c)** the three-dimensional snapshots of the tumor at time points $t = 50$, 88 and 107. Note that blue color represents proliferating cells, red represents migrating, green represents quiescent and grey represents dead tumor cells; one 'time step' is equal to 2.5 hours.

**Figure 4. (a)-(c)**

While at an early stage (**Figure 4(a)**) the proliferate tumor core appears to be completely surrounded by a cloud of migratory cells, at a later time point, a more heterogeneous picture emerges (**Figure 4(b)**). Ultimately, a tip-population of migratory cells can be found adjacent to the location of the source in cube 4 (**Figure 4(c)**; compare also with **Figure 1**).

**Phenotypic behavior:** Not surprisingly, due to the comparably low molecular values of the cells' triggering network, at early time points overall tumor growth is largely dominated by the proliferative phenotype (**Figure 5**). However, when molecular values increase sufficiently, more cells switch (after $t = 41$) to the migratory trait. While the gains in the proliferative cell population overall appear to correspond well to the increases seen in the tumor system's volume curve (**Figure 3**, *above*), the oscillatory behavior in these phenotypic sub-population dynamics suggests that after $t = 52$, an intermediate, migration-*dominant* expansion phase triggers a second proliferation-*dominant* growth phase after $t = 81$.

**Figure 5.**

**Molecular phenotype-switching profiles:** We have also investigated the percentage change in the components of the EGFR network[1] and thus the molecular profile that leads to the phenotypic switch. We focused on time points, $t = 52$ and $81$, where the population curves cross or nearly cross, i.e., $t = 116$ (compare with **Figure 5** (*above*)). While it is evident from the 'migration-to-proliferation' switch that the *qualifying* molecular profile remains very similar over time (**Figures 6.(d)** and **6.(e)**) the situation becomes less stationary in the 'proliferation-to-migration' events (**Figures 6(a)-(c)**). For instance, while a percentage increase in $PLC_\gamma$ active phosphorylated Ca-bound ($X_{11}$) seems to remain a requirement for the switch towards migration throughout, $PLC_\gamma$ inactive Ca-bound ( $X_{10}$ ) appears to start playing a more significant role at later stages of tumor expansion ( $t = 116$ ).

---

[1] Note: cell cycle components are not listed as they are currently only activated once a cell decides to proceed to proliferation.

**Figure 6.(a)-(e)**

## 5. DISCUSSION & CONCLUSIONS

For a variety of cancers, the epidermal growth factor receptor, EGFR, has been shown to be

critically involved in directional motility, or chemotaxis, both *in vitro* and *in vivo* (Bailly et

al., 2000; Soon et al., 2005; Wyckoff et al., 2000). However, there have been conflicting

reports in the literature regarding the prognostic significance of EGFR gene amplification and

over-expression in patients with high-grade gliomas, particularly glioblastomas (for a recent

review, see e.g. Quan et al., 2005). It then becomes apparent that crucial biological

information is being lost by focusing on the gene level *only*, without consideration of EGFR's

extensive protein downstream signaling cascade and its influence on the cell's phenotypic

behavior, respectively. Ideally, one would want to investigate the dynamics of the subcellular

*gene-protein interaction network* without losing sight of the multicellular patterns and ultimately clinical prognosis any such molecular events may lead to. While it is difficult if not impossible, at least for the moment, to realize this in a single experimental setup, integrative computational modeling is advancing towards a level that can provide valuable insights. Here, we present such a novel 3D multi-scale agent-based model that encompasses the macroscopic, microscopic and molecular scale of a virtual brain tumor. Each *in silico* cell is equipped with an EGFR gene-protein interaction module that connects to a simplified cell cycle description.

A tumor cell's phenotypic decision to either proliferate or migrate (or to become quiescent) is determined by the dynamic change in the values of the sub-cellular network's molecular components. These concentration values are impacted by cell-cell signaling, such as through $TGF_\alpha$, and by environmental cues, including nutrients such as glucose, and hypoxia.

The results show that over time proliferative and migratory cell populations not only *oscillate*, thus suggesting a dynamic relationship, rather, this microscopic behavior directly impacts also the spatio-temporal expansion of the entire cancer system. That is, the secondary increase in the proliferative cell population seen after time step 81 (**Figure 5**) is paralleled by a substantial acceleration of the spatio-temporal expansion of the entire tumor system (**Figure 3**). Most intriguingly, however, the percentage change in the concentration of the network's molecular components varies, in some instances considerably. A specific example is inactive Ca-bound $PLC_\gamma$ ($X_{10}$), a network component that appears to start playing a more significant role for the transition to a migratory phenotype at later stages of tumor expansion (**Figure 6c**). In fact, the behavior of the two phenotypic cell populations over time warrants a more detailed inspection: while the first 'crossover' (**Figure 5**), i.e., the increase in the migratory cell population over the proliferative population at time step 52, can be easily explained by the gradual increase of active $PLC_\gamma$ ($X_{11}$), yielding a higher migratory potential according to **Eq. 21**, the second crossover at time step 81 is unexpected since the concentration of active $PLC_\gamma$ ($X_{11}$) continues to build up. One can argue that this is an *emergent* property of the system since no *a priori* condition in the algorithm force such behavior. Afterwards, the migratory cell fraction rapidly re-approaches the proliferative population in time step 116. We note here that since many new cells are generated around time step 107 and since (in this first iteration) a genetically stabile progeny inherits the molecular values of the parental cells, these daughter cells are equipped with an already high concentration of active $PLC_\lambda$ ($X_{11}$) that predisposes them (through **Eq. 21**) to a rapid transition towards the migratory phenotype, much like we saw in the earlier steps of tumor growth.

Aside from its technical merits, admittedly, our modeling approach relies on several simplifications on the biology side and thus inevitably harbors a number of drawbacks. For instance, currently, (apart from the cell-cycle module) the underlying EGFR network itself operates with only two genes, *EGFR* and *TGF*$_\alpha$, and as such a variety of related signaling pathways, such as the mitogen activated protein kinase (MAPK) cascade (Schoeberl et al., 2002) and others, deserve proper consideration in future iterations of the model. Also, for the moment, the algorithm is focused on a set of epigenetic changes and does not yet take heterogeneity inducing genetic instability, a hallmark of cancer progression, into account. As such, the current setting does not yet allow for dynamic alterations of EGFR gene and protein during tumorigenesis, such as through amplification and over-expression as reported widely for high grade gliomas (e.g., Ekstrand et al., 1991). Efforts are, however, underway in our group to address some of these shortcoming in future works.

Nonetheless, our results already confirm the impact post-translational regulation can have on tumor cell behavior, both on the single cell and multicellular level. Furthermore, indicating that over time, differing molecular network states may be able to trigger similar phenotypic behavior, our findings question the value of single time point gene-expression assessments for clinical predictions, thus corroborating recent reports from Rich et al. (2005) and Ouan et al. (2005) that find no prognostic value in EGFR gene amplification. Rather, our results postulate *dynamic* monitoring of a tumor's gene-protein interaction level with techniques such as phospho-proteomics (Blagoev et al., 2003) that already have demonstrated their value as outcome predictors for patients with brain tumors (Schwartz et al., 2005). Absent any clinically available non-invasive molecular imaging, monitoring of the spatio-temporal dynamics in gene-protein interaction levels would have to be achieved with specimen from consecutive biopsies, in parallel to assessing the tumor system's overall expansion through MR imaging time series. Intriguingly, a recent clinical study on such 'image-guided proteomics' seems to support our findings at least in part as these authors report distinctively different protein expression profiles in the glioblastomas' contrast-enhanced rim zone (which is arguably where the 'proliferation-to-migration' (**Figure 6** (**a**)–(**c**)) transition occurs) despite similar histological findings (Hobbs et al., 2003).

In conclusion, this novel three-dimensional computational model is an important step in simulating tumor growth dynamics over multiple scales of interest. While extensions will be necessary to account in greater detail for the complexity of the biology involved, we believe that if properly combined with experimental data, advanced *in silico* platforms such as this one will evolve into powerful integrative research platforms that improve our understanding of tumorigenesis.

# ACKNOWLEDGEMENTS

This work has been supported in part by NIH grants CA 085139 and CA 113004 and by the Harvard-MIT (HST) Athinoula A. Martinos Center for Biomedical Imaging and the Department of Radiology at Massachusetts General Hospital. We thank Dr. Costas Strouthos (Complex Biosystems Modeling Laboratory, Massachusetts General Hospital) for critical review of the manuscript.

## CAPTIONS

**Figure 1:** Shown is the underlying 3D lattice, with cube 4 harboring the nutrient source (see text for details).

**Figure 2:** The diagram displays the sub-cellular EGFR gene-protein interaction network that combines nucleus, cytoplasm and membrane compartments.

**Figure 3:** Depicted is the tumor system volume (*y-axis*) over time (*x-axis*). Shown are mean values of 10 simulation runs with random standard deviation of the glucose concentration at a range between 50 and 60 mmol/l.

**Figure 4:** Shown are 3D snapshots of the tumor system at time step $t = 50$ (**a**), $t = 88$ (**b**), $t = 107$ (**c**).

**Figure 5:** Shown are population dynamics (*y-axis*) of the proliferative (*closed circles*) and migratory cell populations (*open circles*) over time (*x-axis*).

**Figure 6:** Depicted are the average molecular "proliferation-to-migration" profiles (of $n$ cells) at time step (**a**) $t = 52$ ($n = 3$), (**b**) $t = 81$ ($n = 43$), and (**c**) $t = 116$ ($n = 161$), as well as the average "migration-to-proliferation" profiles at time step (**d**) $t = 81$ ($n = 485$), and (**e**) $t = 116$ ($n = 114$). The *x-axis* denotes the network components (from $X_1$ to $X_{14}$; compare with **Fig. 2**), while the *y-axis* represents the percentage change [%] of these molecular species.

**Table 1:** Symbols of the EGFR gene-protein interaction network taken from the literature (Athale et al., 2004; Thorne et al., 2004; Sander et al., 2002)[*].

**Table 2:** Coefficients of the EGFR gene-protein interaction network taken from the literature (Athale et al., 2004; Thorne et al., 2004; Sander et al., 2002).

**Table 3:** Symbols of the cell cycle module taken from the literature (Tyson and Novak 2001; Alarcon et al., 2004).

**Table 4:** Coefficients of the cell cycle module taken from the literature (Tyson and Novak 2001; Alarcon et al., 2004).

---

[*] Note (for **Tables 1-4**): Reasonable estimates were used where no published values were available.

# FIGURES & TABLES

**FIGURE 1.**

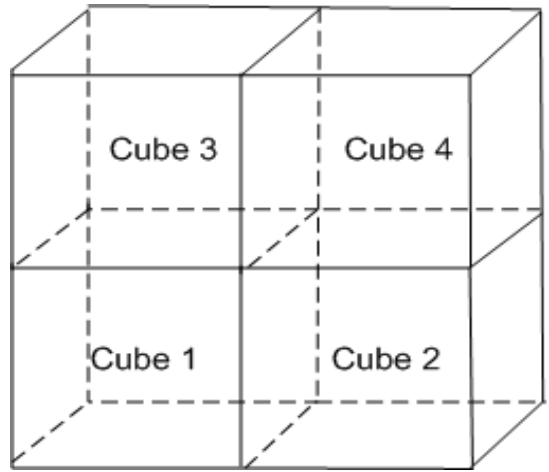

**FIGURE 2.**

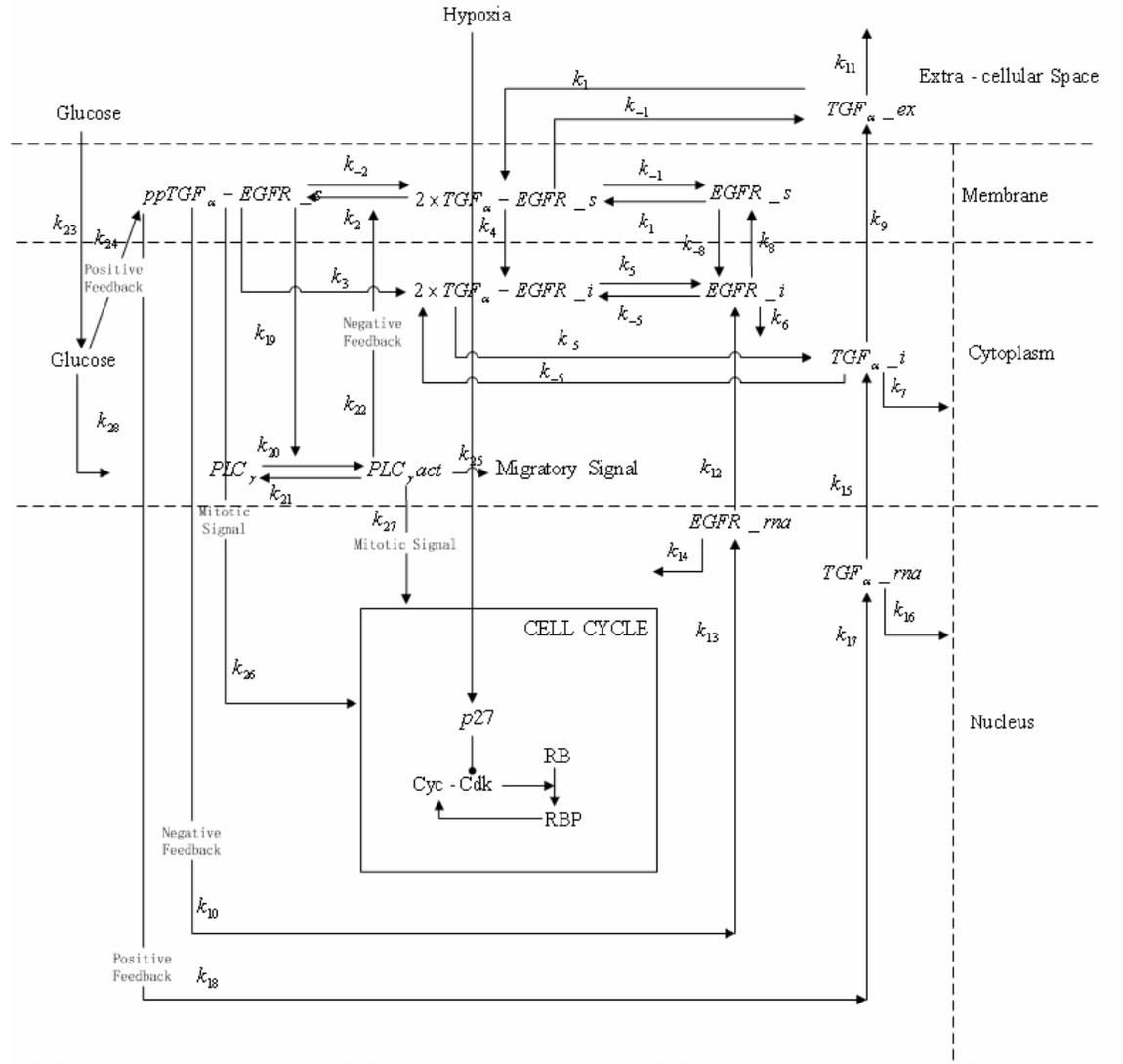

**FIGURE 3.**

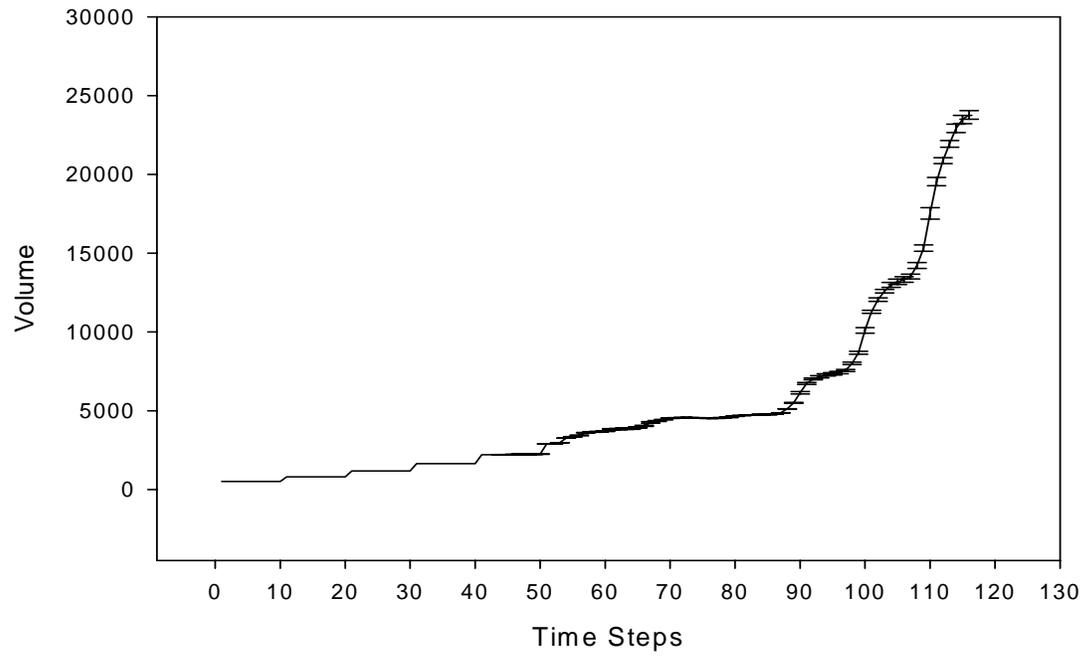

**FIGURE 4.(a)**

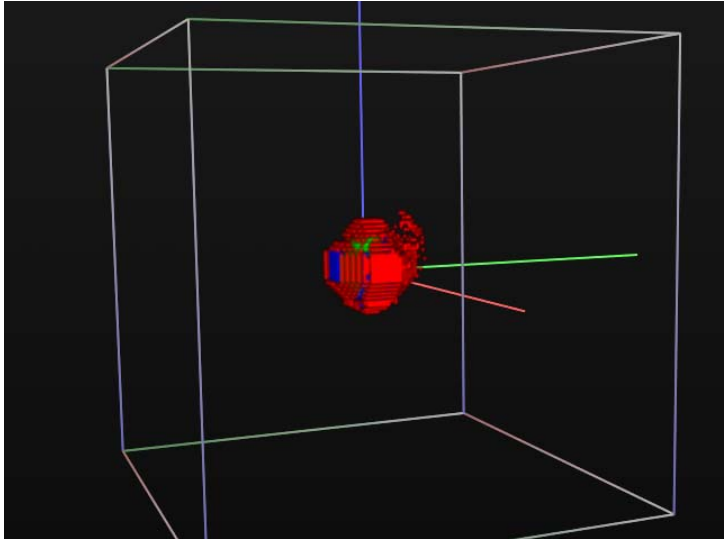

**FIGURE 4.(b)**

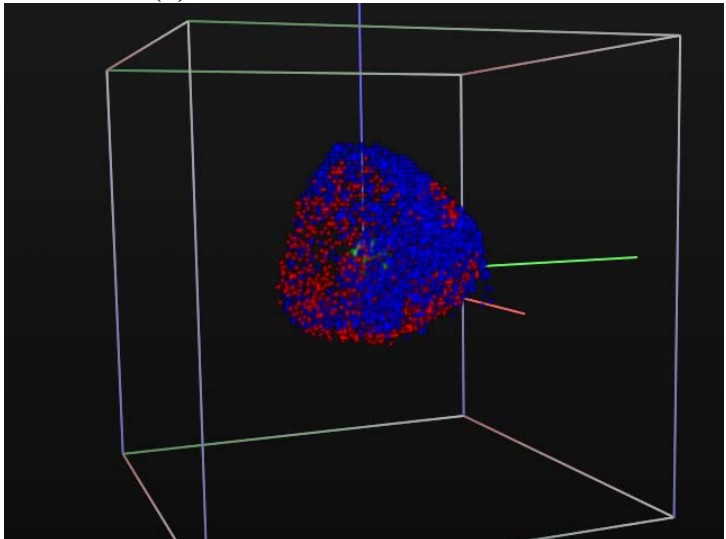

**FIGURE 4.(c)**

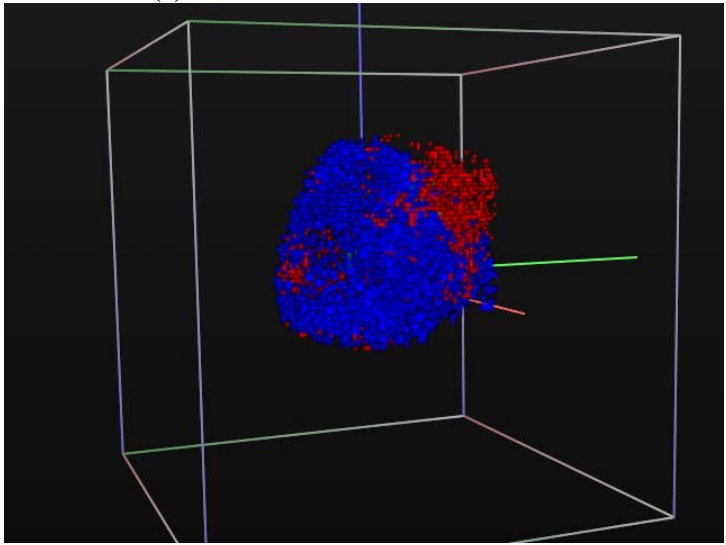

**FIGURE 5.**

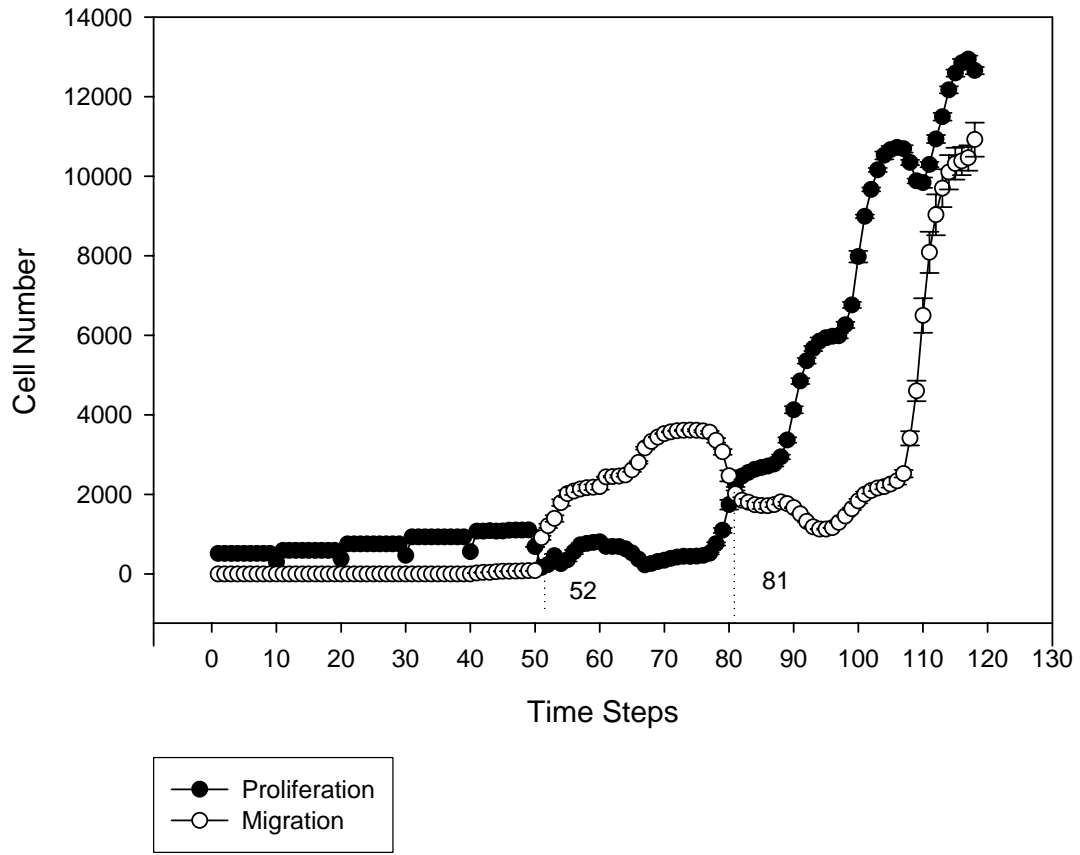

**FIGURE 6.(a)**

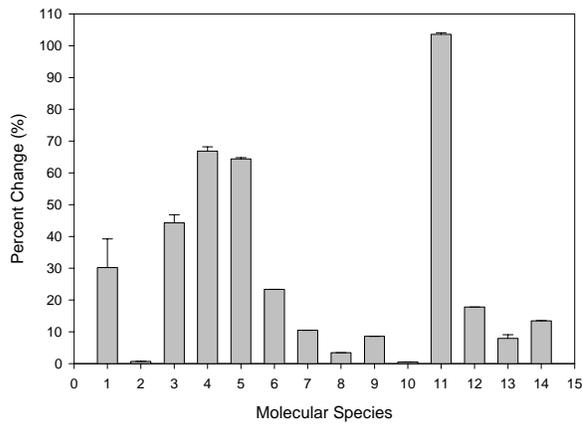

**FIGURE 6.(b)**

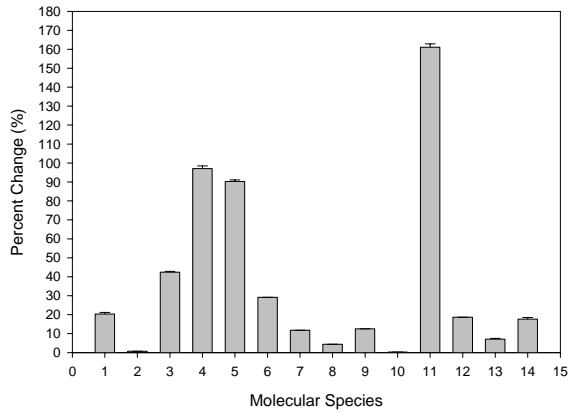

**FIGURE 6.(c)**

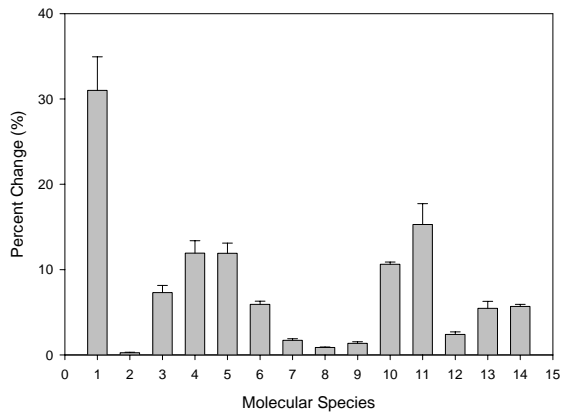

**FIGURE 6.(d)**

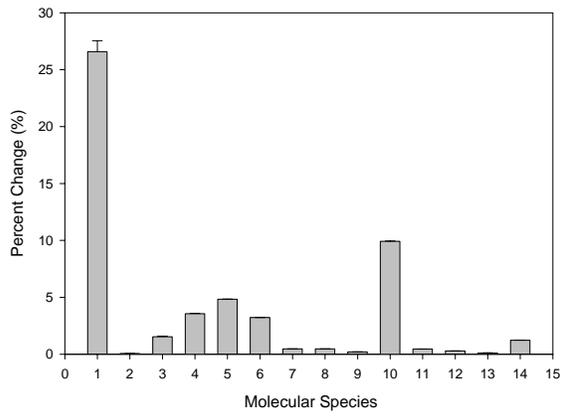

**FIGURE 6.(e)**

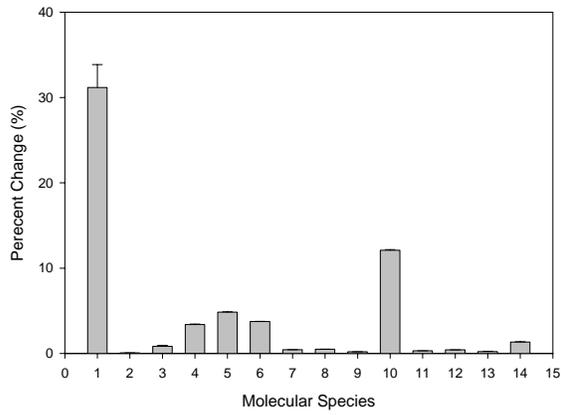

**Table 1.**

| Symbol | Variable | Initial Value [ $nM$ ] |
|---|---|---|
| $X_1$ | $TGF\alpha$ extracellular protein | 1 |
| $X_2$ | $EGFR$ cell surface receptor | 25 |
| $X_3$ | Dimeric $TGF\alpha - EGFR$ cell surface complex | 0 |
| $X_4$ | Phosphorylated active dimeric $TGF\alpha - EGFR$ cell surface complex | 0 |
| $X_5$ | Cytoplasmic inactive dimeric $TGF\alpha - EGFR$ complex | 0 |
| $X_6$ | Cytoplasmic $EGFR$ protein | 0 |
| $X_7$ | Cytoplasmic $TGF\alpha$ protein | 1 |
| $X_8$ | $EGFR$ $RNA$ | 1 |
| $X_9$ | $TGF\alpha$ $RNA$ | 0 |
| $X_{10}$ | $PLC_\gamma$ inactive , Ca-bound | 1 |
| $X_{11}$ | $PLC_\gamma$ active, phosphorylated, Ca-bound | 0 |
| $X_{12}$ | Nucleotide pool | 5 |
| $X_{13}$ | Glucose cytoplasmic | 1 |
| $X_{14}$ | Glucose extracellular | 0 |

**Table 2.**

| Coefficient | Value | Units | Description |
|---|---|---|---|
| $k_1$ | $3\times10^{-3}$ | $nM^{-1}s^{-1}$ | $TGF\alpha - EGFR$ cell–surface complex formation rate |
| $k_{-1}$ | $3.8\times10^{-3}$ | $s^{-1}$ | Rate of dissociation of $TGF\alpha - EGFR$ cell-surface complex |
| $k_2$ | $1\times10^{-3}$ | $s^{-1}$ | Rate of $TGF\alpha - EGFR$ phosphorylation |
| $k_{-2}$ | $1\times10^{-6}$ | $s^{-1}$ | Rate of $TGF\alpha - EGFR$ dephosphorylation |
| $k_3$ | $5\times10^{-5}$ | $s^{-1}$ | Rate of cell-surface $TGF\alpha - EGFR$ internalization |
| $k_4$ | $5\times10^{-5}$ | $s^{-1}$ | Rate of phosphorylated $TGF\alpha - EGFR$ internalization |
| $k_5$ | $1\times10^{-2}$ | $s^{-1}$ | Dissociation rate of cytoplasmic $TGF\alpha - EGFR$ |
| $k_{-5}$ | $1.4\times10^{-5}$ | $nM^{-1}s^{-1}$ | Reverse dissociation rate of cytoplasmic $TGF\alpha - EGFR$ |
| $k_6$ | $1.67\times10^{-4}$ | $s^{-1}$ | Rate of cytoplasmic $EGFR$ protein degradation |
| $k_7$ | $1.67\times10^{-4}$ | $s^{-1}$ | Rate of cytoplasmic $TGF\alpha$ protein degradation |
| $k_8$ | $5\times10^{-3}$ | $s^{-1}$ | Rate of cytoplasmic $EGFR$ insertion into the membrane |
| $k_{-8}$ | $5\times10^{-5}$ | $s^{-1}$ | Rate of cell-surface $EGFR$ internalization |

| $k_9$ | 1 | $s^{-1}$ | Rate of membrane insertion and secretion of $TGF\alpha$ |
|---|---|---|---|
| $k_{10}$ | 0.01 | $s^{-1}$ | Rate of down-regulation of $EGFR$ expression by the $TGF\alpha - EGFR$ complex |
| $k_{11}$ | 0.01 | $s^{-1}$ | Degradation of extracellular $TGF\alpha$ |
| $k_{12}$ | 0.083 | $s^{-1}$ | Rate of translation of $EGFR\ RNA$ |
| $k_{13}$ | 0.036 | $s^{-1}$ | Basal transcription rate $EGFR\ RNA$ |
| $k_{14}$ | $1.2 \times 10^{-3}$ | $s^{-1}$ | $EGFR\ RNA$ degradation rate |
| $k_{15}$ | 0.083 | $s^{-1}$ | Rate of translation of $TGF\alpha$ |
| $k_{16}$ | 0.02 | $s^{-1}$ | $TGF\alpha\ RNA$ degradation rate |
| $k_{17}$ | 0.2 | $s^{-1}$ | Basal transcription rate $TGF\alpha\_rna$ |
| $k_{18}$ | $k_{M1}, V_{M1}, w_1$ | *Dimensionless constant [DC]* | Induction of $TGF\alpha$ transcription by activated $TGF\alpha - EGFR$ at the cell surface |
| $k_{M1}$ | 1 | $nM$ | Km of $TGF\alpha\ RNA$ transcriptional activation |
| $V_{M1}$ | 5 | $nM^{-1}s^{-1}$ | Rate of $TGF\alpha\ RNA$ transcriptional activation |
| $w_1$ | 1 | $DC$ | Weight of Hills' coefficient of $TGF\alpha\ RNA$ activation |
| $k_{19}$ | 0.1 | $nM^{-1}s^{-1}$ | Enhanced rate of $PLC_\gamma$ activation by $EGFR$ |
| $k_{20}$ | 0.1 | $nM^{-1}s^{-1}$ | Basal rate of activation of $PLC_\gamma$ |
| $k_{21}$ | 0.05 | $s^{-1}$ | Rate of in-activation of $PLC_\gamma$ |
| $k_{22}$ | $k_{M2}, V_{M2}, w_2$ | $DC$ | $PLC_\gamma$ dependent rate of de-phosphorylation of phosphorylated $TGF\alpha - EGFR$ |
| $k_{M2}$ | 5 | $nM$ | Km $PLC_\lambda$ inhibition of phosphorylated surface $TGF\alpha - EGFR$ |
| $V_{M2}$ | 0.25 | $nM^{-1}s^{-1}$ | $PLC_\gamma$ inhibition rate of $TGF\alpha - EGFR$ |
| $w_2$ | 1 | $DC$ | Weight of Hill's coefficient $PLC_\lambda$ inhibition of phosphorylated surface $TGF\alpha - EGFR$ |
| $k_{23}$ | $0.1 \times 10^{-3}$ | $s^{-1}$ | Lumped rate of glucose uptake |
| $k_{24}$ | 0.01 | $nM^{-1}s^{-1}$ | Increased rate of $TGF\alpha - EGFR$ phosphorylation by glucose |
| $w_g$ | 5 | $DC$ | Weight of increase in rate of $TGF\alpha - EGFR$ phosphorylation by glucose |
| $k_{25}$ | $3.5 \times 10^{-3}$ | $nM^{-1}s^{-1}$ | Migratory signal |
| $k_{26}$ | $3.5 \times 10^{-3}$ | $nM^{-1}s^{-1}$ | Mitotic signal I |
| $k_{27}$ | 0.0002 | $nM$ | Mitotic signal II |
| $k_{28}$ | 0.7 | $s^{-1}$ | Cytoplasmic glucose rate of degradation |
| $k_{29}$ | 1 | $nM$ | Constant total $PLC_\gamma$ |
| $r_n$ | 0.7 | $mmol / step$ | Glucose exhausting coefficient |
| $D_1$ | $6.7 \times 10^{-7}$ | $cm^2 s^{-1}$ | Diffusion coefficient of glucose |
| $D_2$ | $5.18 \times 10^{-7}$ | $cm^2 s^{-1}$ | Diffusion coefficient of $TGF_\alpha$ |
| $D_o$ | $8 \times 10^{-5}$ | $cm^2 s^{-1}$ | Diffusion coefficient of oxygen tension |

| $T_m$ | $147 \pm 18$ | $pg / ml$ | Maximum concentration of $TGF_\alpha$ |
|---|---|---|---|
| $G_a$ | 17 | $mmol / L$ | Normal concentration of glucose |
| $G_m$ | 57 | $mmol / L$ | Maximum concentration of glucose |
| $S_T$ | 0.3 | $molecules/min$ | Secretion rate of $TGF_\alpha$ |
| $k_a$ | 0.0017 | $DC$ | Normal concentration of oxygen |
| $k_m$ | 0.0025 | $DC$ | Maximum concentration of oxygen |

**Table 3.**

| Symbol | Variable | Initial value [$DC$] |
|---|---|---|
| $X_{15}$ | cdh1-APC complex | 0.9 |
| $X_{16}$ | cyclin-CDK | 0.01 |
| $X_{17}$ | Mass of the cell | 5 |
| $X_{18}$ | Protein p27 | 0 |
| $X_{19}$ | RBNP | 1 |

**Table 4.**

| Coefficient | Value | Units |
|---|---|---|
| $k_{30}$ | 0.004 | $DC$ |
| $k_{31}$ | 0.05 | $DC$ |
| $k_{32}$ | 10 | $min^{-1}$ |
| $k_{33}$ | 35 | $min^{-1}$ |
| $k_{34}$ | 0.04 | $DC$ |
| $k_{35}$ | 0.04 | $DC$ |
| $k_{36}$ | 0.4 | $min^{-1}$ |
| $k_{37}$ | 1 | $min^{-1}$ |
| $k_{38}$ | 0.25 | $DC$ |
| $k_{39}$ | 0.04 | $min^{-1}$ |
| $k_{40}$ | 0.01 | $min^{-1}$ |
| $k_{41}$ | 10 | $DC$ |
| $k_{42}$ | 0.007 | $DC$ |
| $k_{43}$ | 0.01 | $DC$ |
| $k_{44}$ | 0.0017~0.0025 | $DC$ |
| $k_{45}$ | 0.01 | $DC$ |
| $k_{46}$ | 0.01 | $DC$ |
| $k_{47}$ | 0.1 | DC |